\documentclass[conference]{IEEEtran}
\usepackage[square,numbers]{natbib}
\usepackage{hyperref}
\ifCLASSINFOpdf
  \usepackage{graphicx}
  \graphicspath{{./images/}}
\else
\fi
\usepackage{amsmath}
\usepackage{algorithmic}

\begin{document}
\title{Applications of Deep Learning for Ill-Posed Inverse Problems Within Optical Tomography}
\author{\IEEEauthorblockN{Adam N. Peace}
\IEEEauthorblockA{Department of Computer Science\\
University College London\\
London, WC1 3BT \\
Email: \texttt{a.peace@cs.ucl.ac.uk}}}
\maketitle

\begin{abstract}
Increasingly in medical imaging has emerged an issue surrounding the reconstruction of noisy images from raw measurement data. Where the forward problem is the generation of raw measurement data from a ground truth image, the inverse problem is the reconstruction of those images from the measurement data. In most cases with medical imaging, classical inverse Radon transforms, such as an inverse Fourier transform for MRI, work well for recovering clean images from the measured data. Unfortunately in the case of X-Ray CT, where undersampled data is very common, more than this is needed to resolve faithful and usable images.

In this paper, we explore the history of classical methods for solving the inverse problem for X-Ray CT, followed by an analysis of the state of the art methods that utilize supervised deep learning. Finally, we will provide some possible avenues for research in the future.
\end{abstract}

\IEEEpeerreviewmaketitle

\section{Introduction}

\label{sec:intro}

X-Ray Computed Tomography (CT) is a technology that has revolutionized the way in which many fields, including medical imaging, have been able to investigate the inner tomography of bodies in a non-intrusive way, while still allowing the user to overcome the limit of radiography, where all sense of depth is made invisible.

\begin{figure*}
  \center
  \includegraphics[width=\textwidth]{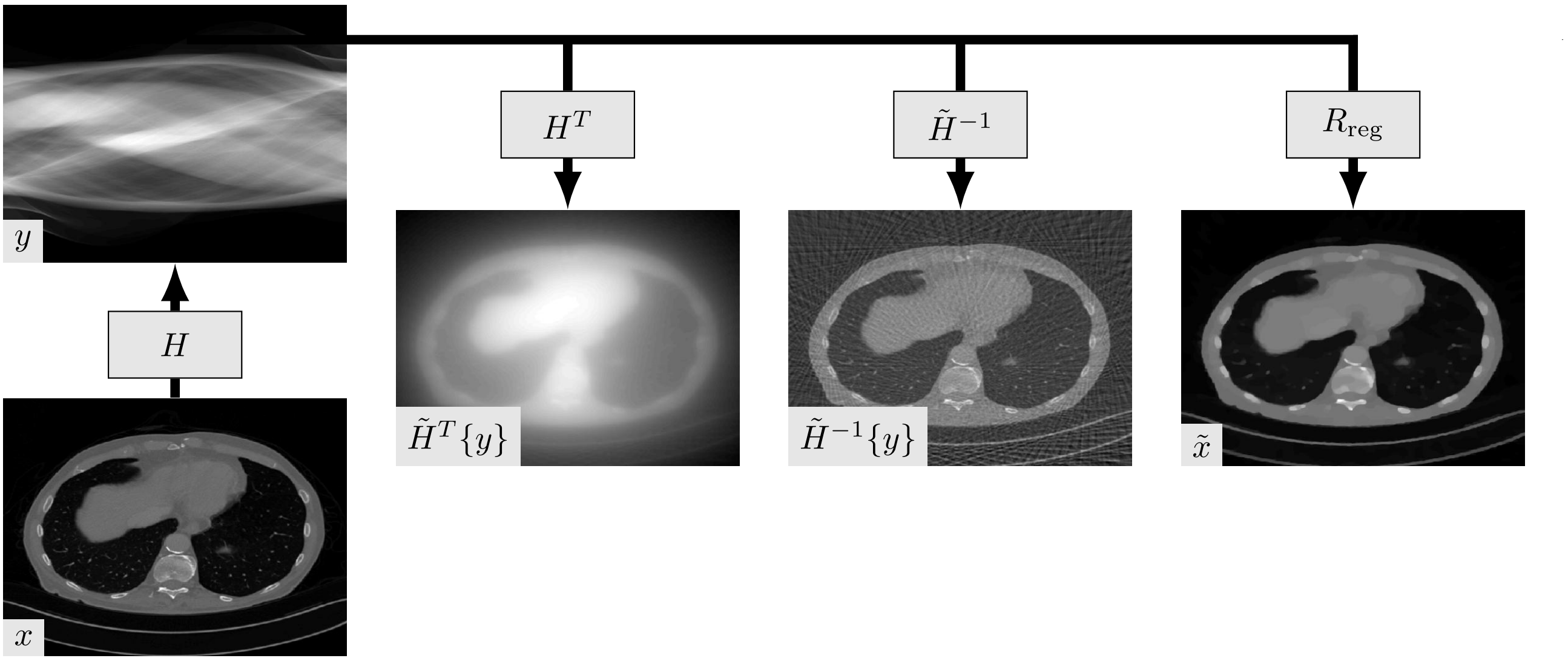}
  \caption{$x \in X$ is true image, $y \in Y$ is the sinogram generated from X-Ray CT scan. The right three images are the result of various reconstruction techniques, with $H^T$ being the back projection, $\tilde{H}^{-1}$ being the filtered back projection where clear artifacts are present and $R_\text{reg}$ is the learning approach with a regularizer, where the image is slightly softened. Adapted from \cite{McCann2017}}
  \label{fig:fbp}
\end{figure*}

The X-Ray CT scan relies on rotating a source and corresponding detector element around the body until scans have been taken at every point in the rotation. These scans are then assembled into a sinogram ($y$ in figure \ref{fig:fbp}), which we call the measurement data. 

One of the most important technologies for X-Ray CT is image reconstruction. To be able to non-intrusively see depth within inner structures in 3 dimensions, it's required to perform some sort of reconstruction function to retrieve original images from measurement data. For perfect situations, where many samples of the same slice are available and the samples are free of noise and artefacts, existing methods work well to reproduce 2D images of each slice. Unfortunately, using an X-Ray source near a human body for long periods of time is unhealthy and can be cancer-inducing. For this reason, we often end up with undersampled data, the effects of which are clear when using classical methods, since all noise in the measurement data is amplified greatly in the reconstruction (As can be seen with $\tilde{H}^{-1}{y}$ in figure \ref{fig:fbp}).

Deep learning is already in the midst of having a transformative impact on so many fields, from playing games better than grandmasters, to computer vision in self-driving cars, to face and speech recognition tasks. \cite{Lecun2015} A recent trend can be seen in the past decade, with deep learning being applied to inverse problems in optical tomography.

In this paper, we aim to give a background of the inverse problem and what it is, followed by a brief history of classical approaches to solving the inverse problem. After that, we will show the current state of the art by investigating the main approaches that have been taken to apply deep learning to image reconstruction and evaluating the papers that have been involved in advancing the field for each approach. We will then deep-dive into the architecture design of a few, key, landmark papers before finally offering a summary of the major challenges faced as well as giving some insight into the future trajectory of the field.


\section{Background}

\subsection{What is the Inverse Problem?}

Formally, the forward problem $H: \mathcal{X} \to \mathcal{Y}$ is a mapping from a true image $x \in \mathcal{X}$ to raw scan data $y \in \mathcal{Y}$ where $H(x) = y$. Although this is a generalized form, with X-ray CT, $x$ represents an image of X-ray attenuations, $H$ represents the physics of the scanner described in section \ref{sec:intro} and $y$ represents the resulting sinogram.

The inverse problem is then the recovery of the original image $x$ from the raw scan data $y$ using a reconstruction function $R: \mathcal{Y} \to \mathcal{X}$ as one can see in fig. \ref{fig:fbp}.

\label{sec:background:size}
The obvious solution is to find the mapping $R: \mathcal{Y} \to \mathcal{X}$ from $H$ to produce the true image. The main issue faced with this is, even for a $256 * 256$ image, the inversion is that of a $2^{16} * 2^{16}$ matrix, which has approximately $10^{12}$ elements, a size which, for multiple slices, would be challenging to even store in modern hardware, let alone actually be able to invert. \cite{arridge2019solving}

\subsection{Classical Solutions to the Inverse Problem}
To achieve this reconstruction, we can model the forward problem $H$ and produce an estimate of $x$ by minimizing the loss function between $y$ and the estimated sinogram from using our modelled $H$:
\begin{align}
  R_{obj}(y) = \text{argmin}_{x \in \mathcal{X}} f(H(x), y)
  \label{eq:inv}
\end{align}

This is called the \textit{objective function approach}. By computing the inverse operator $\hat{H}^{-1}$, we trivially have $R_{obj}(y) = \hat{H}^{-1}(y)$.

For X-Ray CT, $\hat{H}^{-1}$ is known as the filtered back projection algorithm. An even more basic algorithm than that is the standard back projection algorithm $H^T : \mathcal{Y} \to \mathcal{X}$ which just simply assembles the raw data back into the form of an image, but as one can see in figure \ref{fig:fbp}, it is not a great representation of the true image. Even the direct inverse, $\hat{H}^{-1}$, sees noise amplification with undersampled, noisy scan data, due to the problem's ill-posedness, meaning multiple non-unique solutions may exist. This is often the case, since we want to minimize exposing patients to harmful cancer-inducing X-ray radiation, resulting in a limited number of samples.

In the case that the inverse problem is ill-posed, a regularizer is often a better approach. This is where we add a regularization function of the original image to the minimization giving
\begin{align}
  R_{reg} = \text{argmin}_{x \in \mathcal{X}} f(H(x), y) + g(x)
\end{align}

where $g$ is the regularization functional. 

The conditional mean approach does act as a regularizer so is well suited for ill-posed inverse problems, but like Maximum A Posteriori, its operator uses integration and so it is limited in scalability. The Bayesian estimator also acts as a regularizer but it involves minimizing over an integration, making it unfit even for small tasks. \cite{arridge2019solving}

Variational regularizers take the idea one step further by aiming to minimize the loss of another regularization function such as a Tikhonov regularizer \cite{Kaltenbacher2011} or more commonly total variation or total generalized variation. \cite{Lunz2018} Up until recently, using TV as a regularization functional for $g$ was the preferred approach for producing denoised reconstructions of the true image.

Apart from the \textit{objective function approach}, we also have the \textit{learning approach}, where we create a set of true images and corresponding raw sinograms ${(x_i, y_i)}_{i=1}^N$. From this, a learned reconstruction algorithm $R_{learn}$ would be found by minimizing over all possible parameters $\Theta$
\begin{align}
  R_{learn} = \text{argmin}_{R_\theta, \theta \in \Theta}
  \sum_{i=1}^N
  f(x_i, R_\theta{y_i}) + g(\theta)
  \label{eq:learned}
\end{align}

with $f$ being an error function like the 2-norm once again and g being a regularizer of the parameters to avoid overfitting. Once the optimal parameter $\hat\theta$ is selected, $R_{learn}$ is ready to be used for reconstruction of new images.

These two methods have been the state of the art up until recently, but both have their limitations. The \textit{learning approach}, for example, needs to have a training set prepared and can still suffer heavily from overfitting, whereas the \textit{objective function approach} suffers from needing a model of the forward problem, cost function and regularizer. \cite{McCann2017}

\section{Reconstruction with Deep Learning}
\label{sec:approaches}
By using a Convolutional Neural Network, we can perform the optimization in \ref{eq:learned} by setting the set of $R_\theta, \theta \in \Theta$ to be a combination of filters, parameterized by the filter weights, over which the optimization occurs.

Following the massive surge of interest into ill-posed inverse problem in the early 2000s, many new approaches emerged and along with the rise in use of deep learning, three in particular were significant improvements on the classical approaches. We now compare these three here and explore their use within various papers.

\subsection{Learned Denoisers}
Following the emergence of deep learning, the first applications to optical tomography inverse problems was through learned post-processors or learned denoisers, as suggested by \citet{wang2016perspective}. For this approach, filtered back propagation is first performed on the sinogram to produce a noisy image, where a neural network is then used to remove the noise induced by the pseudo-inverse step. By limiting the learning stage to just the transformation $X \to X$ of the pseudo-inverted sinogram reconstruction $x \in X$, we greatly reduce the complexity of the task and authors such as \citet{zhao2017few} were among the first to show significant reduction in noise.

\citet{chen2017low} took a comparable approach by building a convolution-deconvolution network using patched-based training to achieve similar results.

Given the relatively trivial nature of this approach, many groups like \citet{jin2017deep} use more generalized methods of image noise reduction such as U-net or, in the case of \citet{kang2017deep}, using U-net on directional wavelets. \citet{Kang2018} then developed this further to recover more texture in the image by using framelet-based denoising with a wavelet residual network \cite{Dong2017}. 

\citet{Ye2018} then further applied classical signal processing methods in his \textit{Deep Convolutional Framelets} approach, further improving performance compared to the previous attempts.

\subsection{Iterative}
With the learned deep learning denoiser approach having been proven, more and more interest lied in the possibility of further involvement of deep learning in the inverse problem. The learned iterative scheme seeks to learn the entire transformation from measurement domain to reconstruction domain. The issue with this, as explained in \ref{sec:background:size}, is the great computational expense of performing the reconstruction process in one step. \cite{adler2018learned}

To overcome this, it was proposed by \citet{Yang2016} to utilise learned iterative schemes that resembled classical optimization techniques but instead used machine learning to perform the optimization by making updates based on the result of applying the forward operator on the previous iteration. In this case, the operation was performed for MRI reconstruction

This idea was then further developed by \citet{Putzky2017} for general applications and then for X-Ray CT by \citet{adler2018learned} who developed the highly performant Primal-Dual reconstruction algorithm.

\subsection{End-to-end}

The most recent advancement in the field, introduced for MRI by \citet{schlemper2017deep} and \citet{hammernik2018learning} and later pioneered for X-Ray CT by \citet{Zhu2018} with their \textit{AUTOMAP} is the end to end approach. Unlike the iterative approach, \textit{AUTOMAP} is a fully learned algorithm that produces a reconstruction of the true image from the measurement. 
\begin{figure}[h]
  \centering
  \includegraphics[width=0.48\textwidth]{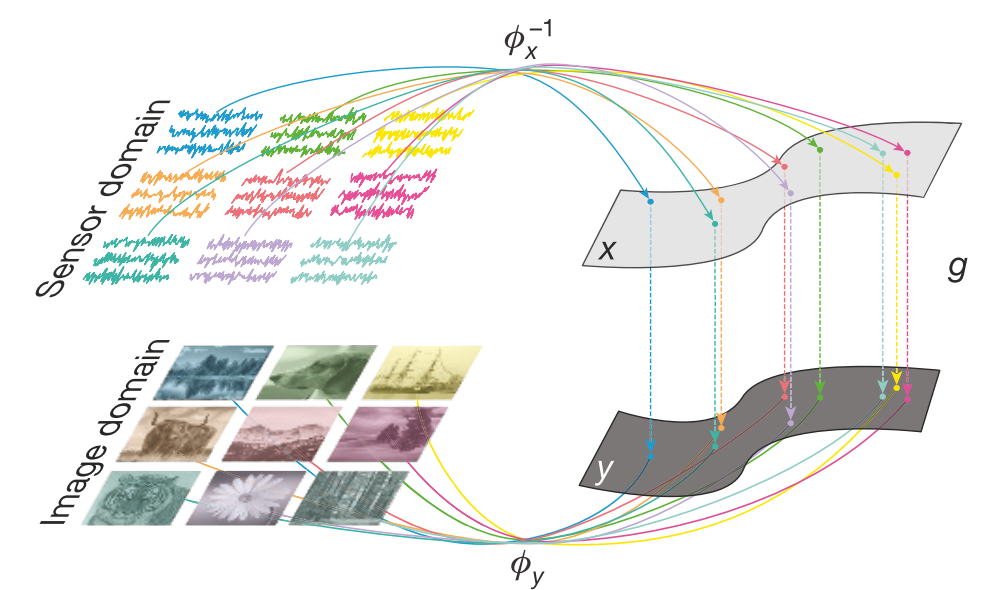}
  \caption{Illustration of the learning performed by \textit{AUTOMAP} to map measurement data to image data. Adapted from \cite{Zhu2018}}
  \label{fig:zhu2018b}
\end{figure}

As is visible in figure \ref{fig:zhu2018b}, a neural network is trained to map raw measurement data to clean images in the training stage. This network can then be applied to new measurement data to produce clean images.

Although as explained before, this is a more computationally expensive method, \textit{AUTOMAP} does successfully learn the entire image reconstruction process for low-resolution images and is more performant than traditional methods.

\section{Architecture}
The deep neural networks used in some of the papers are sensitive to the architecture they employ. In this section, we look at the architectures used in some landmark papers.

\citet{jin2017deep} use a modified U-net convolutional network in their deep learned denoiser. 

\begin{figure}[h]
  \centering
  \includegraphics[width=0.48\textwidth]{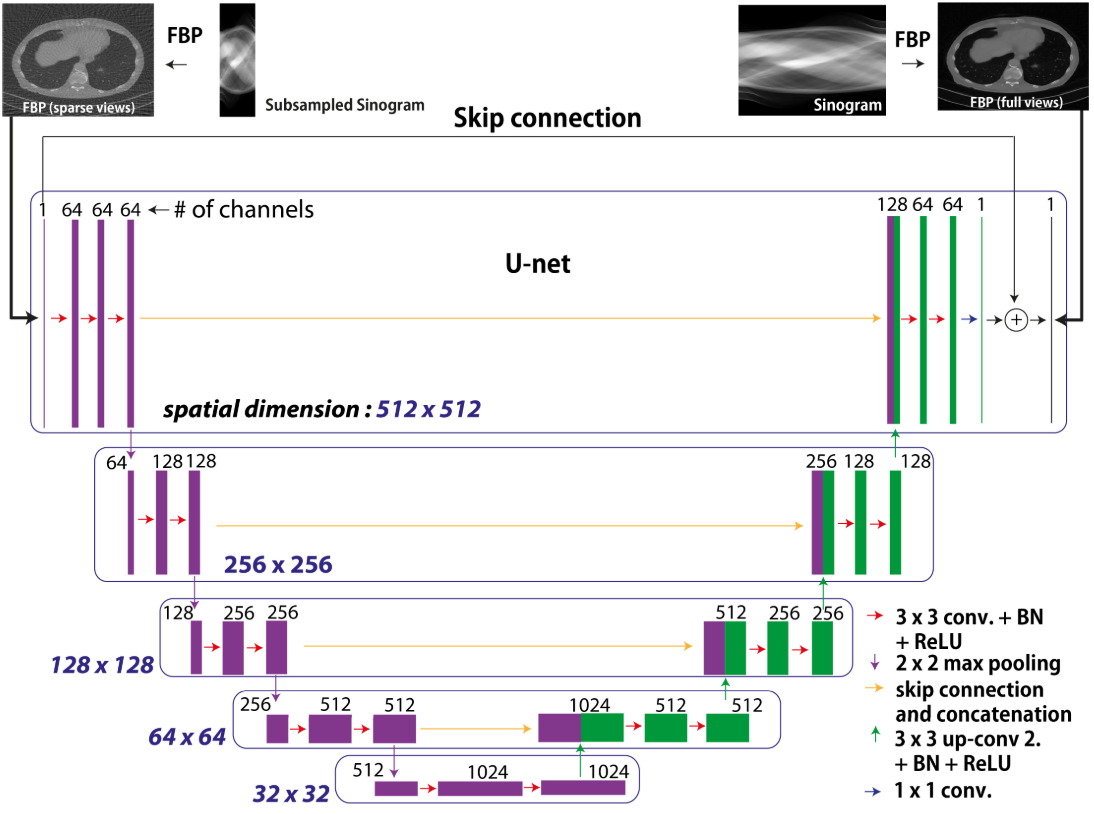}
  \caption{Network architecture used by \citet{jin2017deep}}
  \label{fig:jin2017deepArch}
\end{figure}

By using a dyadic scale decomposition, the filter sizes at layers at the top and bottom are smaller than those in the middle layers. Because of the nature of the forward problem, fixed filter sizes throughout all the layers may not have been sufficient to invert the function effectively. As in most convolutional neural networks, U-net uses multichannel filters, meaning many feature maps exist at each layer of the network. Finally, by including a skip connection, the network is able to learn the differences between input and output images.

As a comparison to an iterative approach, \citet{adler2018learned} use a vastly different network architecture due to there existing a mapping from one domain to another.

\begin{figure}[h]
  \centering
  \includegraphics[width=0.48\textwidth]{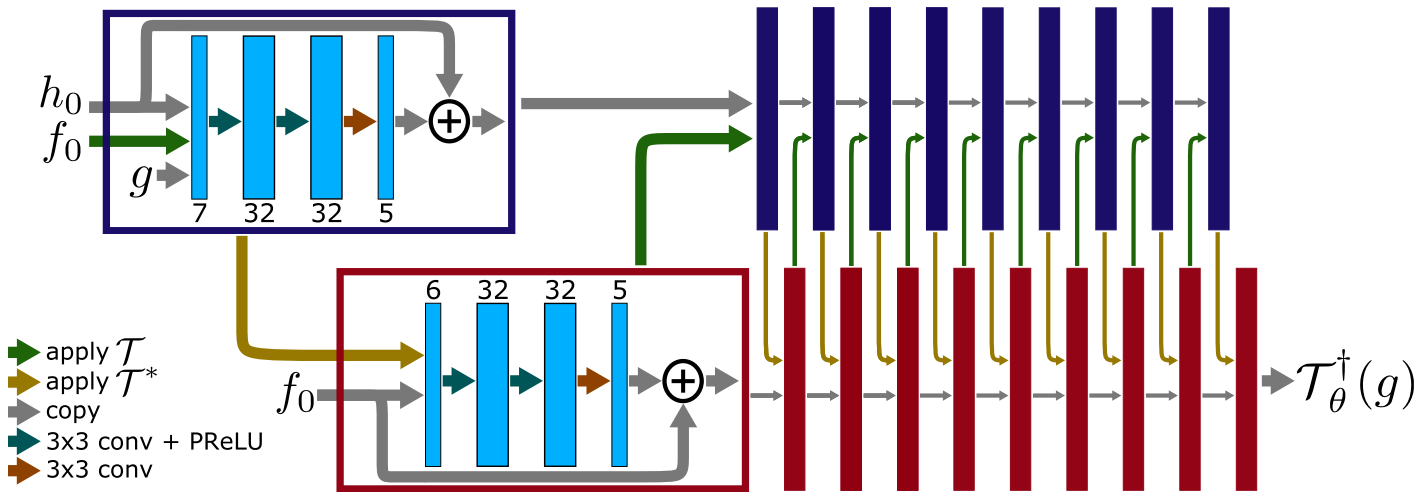}
  \caption{Network architecture used by \citet{adler2018learned} for their \textit{primal-dual} algorithm}
  \label{fig:adler2018learned2}
\end{figure}

In this architecture diagram, the dual iterates are given by the blue boxes and the primal iterates are in the red boxes, with both blue and red boxes having the same architecture. Where this differs from the architecture of a classical \textit{primal dual hybrid gradient} \cite{Adler2017} is that the primal and dual iterates would be given by proximals with over-relaxation parameters as opposed to convolutional neural networks.

Finally, we compare the above two architectures to that of the \textit{AUTOMAP} from \citet{Zhu2018}.

\begin{figure}[h]
  \centering
  \includegraphics[width=0.48\textwidth]{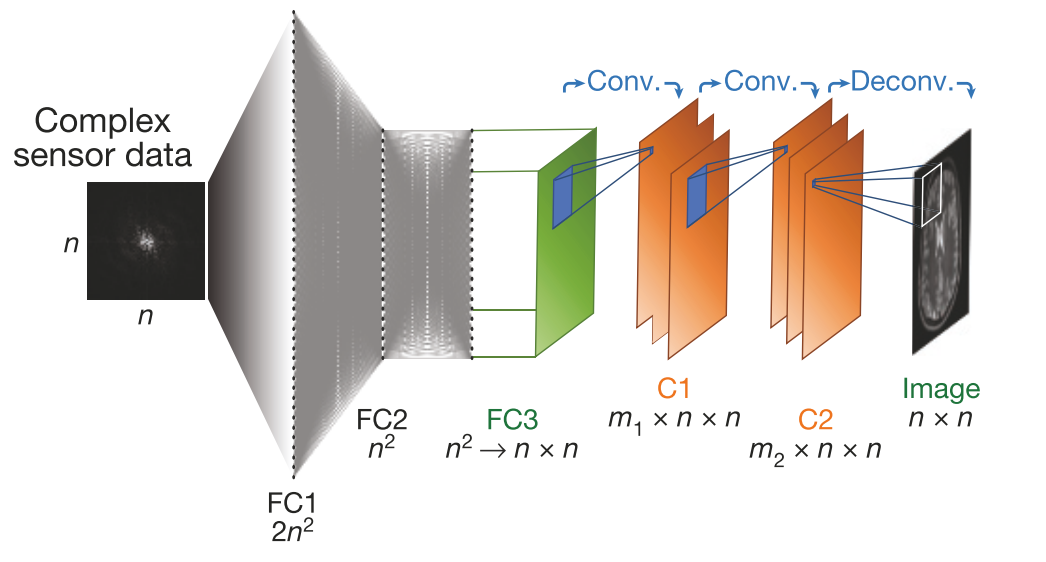}
  \caption{Network architecture used by \citet{Zhu2018} for the \textit{AUTOMAP}}
  \label{fig:zhu2018c}
\end{figure}

With a neural network of fully connected layers, the \textit{AUTOMAP} architecture is interesting because, although the transformation from one domain to another sounds more complicated, fewer convolutional layers are used than in U-Net utilized by \citet{jin2017deep}\footnote{Of note is that the paper by \citet{Zhu2018} is rather challenging to understand so the analysis of the architecture for this paper is limited.}.

\section{Conclusion}
\label{sec:conclusion}

\subsection{Major Challenges}

The major challenges that have been faced in this field include the limitation provided by the harmful nature of the X-Ray CT scan. By using a radioactive source, the number of samples that one is capable of taking is heavily limited. With under-sampled data, the inverse problem becomes ill-posed, making classical approaches either useless or not capable of producing faithful reproductions of the true image.

To overcome this, three approaches have stood out. Firstly, learned denoisers use filtered back projection to produce an image in the domain of the true image then roll out general algorithms such as U-Net to learn a denoising function that, along with filtered back projection, can be applied to new scans. Secondly, the iterative approach learns the transformation from measurement to reconstruction by making updates iteratively. Finally, the latest advancement is surrounding the end-to-end approach, where \citet{Zhu2018} were able to effectively create a fully-learned approach to reconstruct the true image from measurement data.

\subsection{Possible Next Steps}
\begin{enumerate}
  \item \textit{End-to-end Improvements:} With the end-to-end approach having the greatest potential for improvement in the future, we would like to see this approach be developed further. The greatest drawback it faces at the moment is in performance. \textit{AUTOMAP} does struggle heavily with higher-resolution images due to the computational cost involved. Potentially, it would be possible to further combine methods used in the iterative approach with those used in \textit{AUTOMAP}, to create a more effective solution for end-to-end image reconstruction.
  \item \textit{Further Applications:} With many methods being proven at this stage for MRI and for X-Ray CT scans, it would be interesting to see the same theory be applied to other imaging devices such as ultrasound, microscopy and PET. This advancement could help improve what is possible across the whole field of medical imaging.
  \item \textit{Generalized Approaches:} We would like to see the recent research into transfer learning \cite{Espeholt2018} used for solving the inverse problem in the the real world where variables like noise levels and image resolutions change regularly.
\end{enumerate}

\section*{Acknowledgment}

\textit{A special thanks to Simon Arridge for kindly offering his assistance in performing the research for this paper.}

\newpage
\IEEEtriggercmd{\enlargethispage{-5in}}


\bibliographystyle{unsrtnat}
\bibliography{library}

\end{document}